\documentclass[12pt,a4paper]{article}
\usepackage{graphicx}
\usepackage{amsmath}
\usepackage{amssymb}
\usepackage{cite}
\usepackage[height=8.8in,width=6.45in]{geometry}
\usepackage{caption}
\begin{document}
\begin{titlepage}
\hfill CALT-TH 2017-21, IPMU 17-0071\\
\vbox{
    \halign{#\hfil         \cr
           } 
      }  
\vspace*{17mm}
\begin{center}
{\Large \bf On Distinguishability of Black Hole Microstates}

\vspace*{15mm}

{\large Ning Bao${}^{1,2}$ and Hirosi Ooguri${}^{1, 3}$}

\vspace*{8mm}

${}^1$ Walter Burke Institute for Theoretical Physics \\ California
Institute of Technology,
 Pasadena, CA 91125, USA\\
\vspace*{0.3cm}
${}^2$ Institute of Quantum Information and Matter \\ California
Institute of Technology,
 Pasadena, CA 91125, USA\\
\vspace*{0.3cm}
${}^3$ Kavli Institute for the Physics and Mathematics of the Universe
\\ University of Tokyo,
Kashiwa, 277-8583, Japan

\vspace*{9mm}


\end{center}
\begin{abstract}
We use the Holevo information to estimate distinguishability of microstates of a
black hole in anti-de Sitter space by measurements one can perform on a subregion of a
Cauchy surface of the dual conformal field theory. 
We find that microstates are not distinguishable at all
until the subregion reaches a certain size and that
perfect distinguishability can be achieved before
the subregion covers the entire Cauchy surface.
We will compare our results with expectations from the
entanglement wedge reconstruction, 
tensor network models, and the bit threads interpretation of
the Ryu-Takayanagi formula.

\end{abstract}

\end{titlepage}

\vskip 1cm
\section{Introduction}

In the AdS$_{d+1}$/CFT$_d$ correspondence\cite{Maldacena:1997re}, the eternal black hole 
of inverse temperature $\beta$ described by the anti-de Sitter (AdS) Schwarzschild 
solution in $(d+1)$ dimensions is dual to the thermal ensemble, 
\begin{align}
  \rho = \sum_i p_i \rho_i,
  \label{thermal}
  \end{align}
  where
 \begin{align}
  p_i = \frac{e^{-\beta E_i}}{\sum_i  e^{-\beta E_i}},
 \end{align}
 and  $
  \rho_i = |\psi_i \rangle \langle  \psi_i |$'s are orthonormal enegy eigenstates in the Hilbert space ${\cal H}$
 of the conformal field theory (CFT) in $d$ dimensions. We assume that the ensemble is in the
 high temperature phase $\beta^{-1} > (d-1)/2\pi$ so that
 the bulk geometry is dominated by the AdS-Schwarzschild solution \cite{Hawking:1982dh}.
 
Consider one of the microstates $\rho_i$ in some small energy-band around the mass of the black hole,
which is the average energy of the ensemble. 
By the eigenstate thermalization hypothesis \cite{Deutsch, SrednickiETH}, we expect that 
it is described by a bulk geometry approximately equal to the 
AdS-Schwarzschild solution, in the sense that 
expectation values of a certain set of observables that probe 
 outside of the
horizon can be evaluated using the solution\footnote{We assume 
 that the black hole is
spherically symmetric and at the center of AdS. Thus, we restrict each $\rho_i$ in the sum (\ref{thermal}) to have zero angular momentum. 
 If  CFT has additional global symmetry, we assume that $\rho_i$ is in its trivial representation.}.
On the other hand, $\rho$ and $\rho_i$ are clearly different states.
If we are allowed to make arbitrary measurements on CFT, we should be able to distinguish between them. 

The purpose of this paper is to discuss to what degree one can differentiate one microstate from another  
under a restricted set of measurements. We will consider measurements that can be performed 
on a subregion $A$ of a Cauchy surface $S^{d-1}$ of the CFT on $\mathbb{R}_{{\rm time}} \times S^{d-1}$, which is
the boundary of AdS. We will estimate distinguishability between microstates by using the Holevo information, which gives 
the upper bound of the mutual information between any measurement on
 the region $A$ and microstates \cite{Hol}.
 We find that microstates are not distinguishable at all
until the subregion reaches a certain size and that perfect distinguishability can be achieved before
the subregion covers the entire Cauchy surface.
We will compare our results with expectations from the
entanglement wedge reconstruction  \cite{Headrick:2014cta, Dong:2016eik, Cotler:2017erl}, 
tensor network models \cite{Vidal:2007hda, Swingle, Happy, Random},  and the bit threads interpretation  \cite{Freedman:2016zud} of
the Ryu-Takayanagi formula \cite{Ryu:2006bv}.

\begin{figure}[t]
  \begin{center}
    \includegraphics[width=0.6\textwidth]{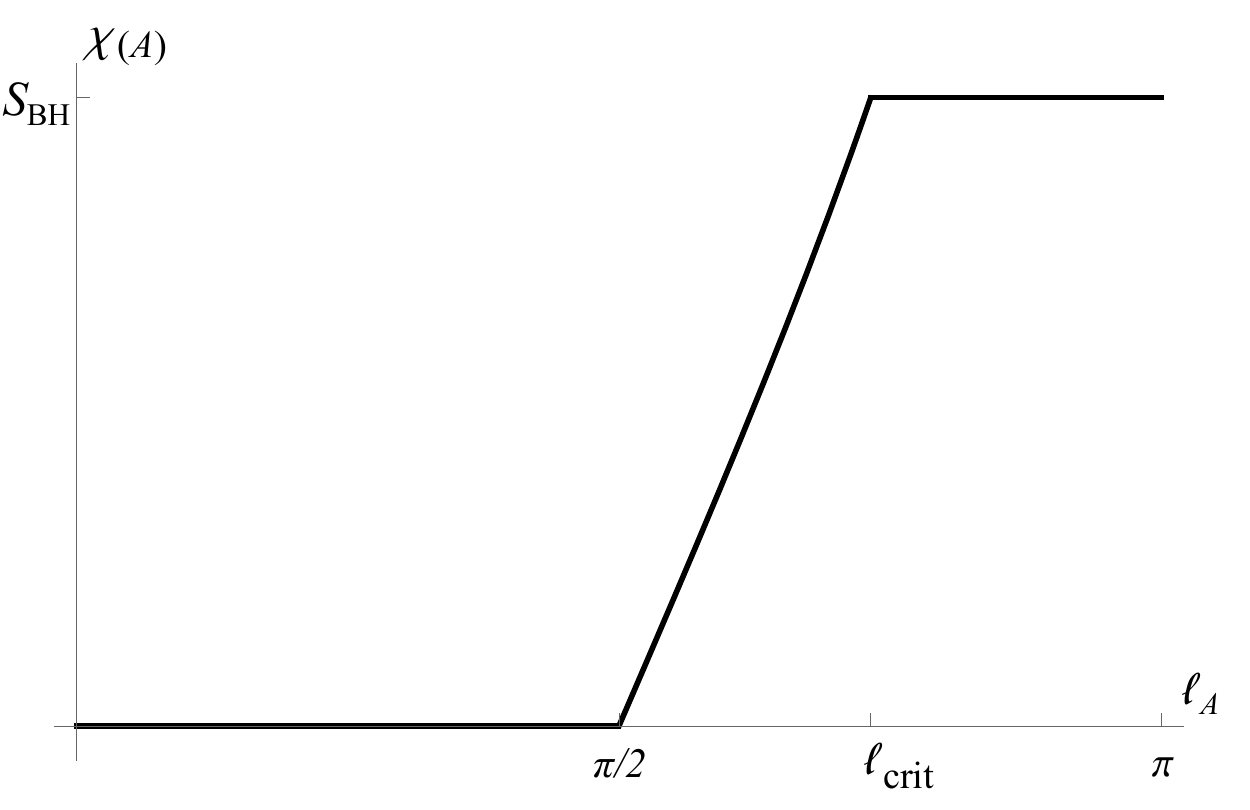}
\end{center}
  \caption{The Holevo information $\chi(A)$ in AdS$_3$/CFT$_2$, as a function of the radius $\ell_A$ of the 
  subregion $A$.
  Note that $\chi(A)$ is identically equal to zero until $A$ covers one half of the Cauchy surface of CFT.
  Above this point, $\chi(A)$ increases monotonically until 
 it achieves the maximum value, which is the Bekenstein-Hawking entropy $S_{{\rm BH}}$ of the black hole,
  at $\ell_{{\rm crit}}$ defined in section 2.}
\end{figure}

This paper is organized as follows. In section 2, we introduce the notion 
of the Holevo information $\chi(A)$ as a measure of distinguishability
between microstates. In sections 3, we compute  $\chi(A)$ in AdS$_3$/CFT$_2$. 
As depicted in Figure 1, 
we find that $\chi(A)$ is identically equal to zero until the subregion $A$ covers one half of
the Cauchy surface of CFT$_2$, showing that there is no mutual information between
microstates and measurements on $A$ in this range. 
As the subregion $A$ becomes larger,
$\chi(A)$ starts to increase monotonically until it reaches a critical size, where $\chi(A)$
achieves the maximum possible value.
We note that this plateaux phenomena in the context of the difference between the entanglement of a CFT boundary region and that of its complement was presented in detail in \cite{Hubeny:2013gta}.
In section 4, we generalize these results to AdS$_{d+1}$/CFT$_d$ for arbitrary $d$. 
 In section 5, we  compare our results with expectations from the pictures of
entanglement wedge reconstruction,
tensor network models,  and the bit threads interpretation of
the holography.
We also point out that one
can reduce the critical size of $A$ where $\chi(A)$ attains
the maximum value, by allowing $A$ to be disjoint.

\section{Holevo Information}

As the first step in quantifying distinguishability under
 measurements on the subregion $A$, we consider the relative entropy, 
\begin{align}
   S(\rho_{i, A} || \rho_A) =  - {\rm tr}
   \left(   \rho_{i, A} \log \rho_{A} \right) + {\rm tr} \left( \rho_{i, A} \log \rho_{i, A} \right),
\end{align}
where $\rho_{i,A}$ and $\rho_A$ are partial traces of $\rho_i$ and $\rho$ over the 
subspace of the CFT Hilbert space associated to the complement of $A$ on the spacelike
section. The relative entropy
is zero if and only if the two density matrices, $\rho_{i,A}$ and $\rho_A$, are identical. 
It turns out that  the average of the relative 
entropy $S(\rho_{i, A} || \rho_A)$ over the ensemble is related to the von Neumann entropies of $\rho_A$ and $\rho_{i, A}$ as, 
\begin{align}
\sum_i p_i S(\rho_{A,i} || \rho_A) & = 
\sum_i p_i \left[  - {\rm tr}
   \left(   \rho_{i, A} \log \rho_{A} \right) + {\rm tr} \left( \rho_{i, A} \log \rho_{i, A} \right)\right] \nonumber\\
   & = S(\rho_A) - \sum_i p_i S(\rho_{i,A}) .
   \label{average}
\end{align}

The combination, 
\begin{align}
 \chi(A) = S(\rho_A) - \sum_i p_i S(\rho_{i,A}),
 \label{Holevo}
 \end{align}
on the right-hand side of (\ref{average}) is known as the Holevo information \cite{Hol}. 
Though $S(\rho_A)$ and $S(\rho_{i, A})$ contain ultraviolet divergences,
they cancel with each other in the Holevo information. In fact, $\chi(A)$ 
is bounded both below and above  as,
\begin{align}
0 \leq \chi(A) \leq S_{{\rm Shannon}} \equiv - \sum p_i \log p_i .
\end{align}
The lower bound is due to the concavity of the von Neumann entropies, and
the upper bound  by the Shannon entropy $S_{{\rm Shannon}}$
is satulated if and only if $\rho_{i, A}$ and $\rho_{j, A}$ have orthogonal support for every
pair of $i$ and $j$. 
Therefore, we can use $\chi(A)$ to quantify distinguishability 
of black hole microstates by measurements performed on $A$. 
Given a specific choice of a microstate $\rho_X$, the Holevo information is known to be the upper bound of the mutual information (also known as the "accessible information") between any measurement on
 the region $A$ and identifying information regarding $\rho_X$ \cite{Hol}.

In preparation for our application of the Holevo information to
the black hole in AdS, 
let us make a few simple observations.
When $A$ is the entire Cauchy surface of the CFT, we have
$\rho_A = \rho$, and its von Neumann entropy is equal to the Shannon entropy, which in this case is the Bekenstein-Hawking entropy
$ S_{{\rm BH}}$,
\begin{align}
  S(\rho)  = - \sum_i p_i \log p_i = S_{{\rm BH}}.
  \label{upperbound}
  \end{align}
  On the other hand $S(\rho_i)=0$ since the $\rho_i$'s are pure states. 
  Therefore, $\chi(A) = S_{{\rm BH}}$ in this limit.
  This is as expected since 
  $\rho_i$'s are orthonormal and perfectly distinguishable from each other. 
In the opposite limit where $A$ is infinitesimal compared to the total area of
the spacelike section, both density matrices $\rho_A$ and $\rho_{i, A}$ approach the 
identity, and 
$\chi(A)=S(\rho_A) - \sum_i p_i S(\rho_{i, A}) \rightarrow 0$.  
Since the Holevo information is monotonic in $A$, 
microstates become more distinguishable as $A$ become
larger and attains maximal distinguishability when $A$ covers the entire spacelike region of CFT.

\begin{figure}[t]
  \begin{center}
    \includegraphics[width=0.8\textwidth]{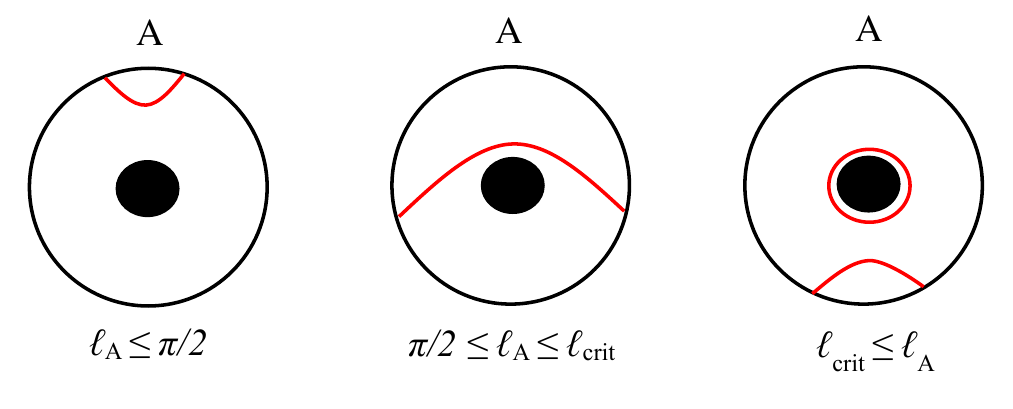}
\end{center}
  \caption{The Ryu-Takayanagi surfaces for region $A$ for the thermal ensemble $\rho = \sum_i p_i \rho_i$.}
\end{figure}

In the large $N$ limit, the entanglement entropy $S(\rho_A)$ for the thermal ensemble 
can be evaluated using the Ryu-Takayanagi formula
 in the AdS-Schwarzschild geometry.
We assume that the black hole is at the center 
of AdS and that the subregion $A$ is ball-like on the Cauchy surface $S^{d-1}$ of 
CFT. We normalize the
radius $\ell_A$ of $A$ so that $\ell_A=\pi/2$ is when one half of the Cauchy surface $S^{d-1}$ is covered by $A$, and 
$\ell_A = \pi$ is when the entire $S^{d-1}$ is covered. 
Since the black hole is rotationally symmetric, the entanglement entropy $S(\rho_A)$ depends only on
$\ell_A$ and the temperature $\beta^{-1}$ of the black hole. Thus we write, 
\begin{align}
S_{{\rm thermal}}(\ell_A) \equiv S(\rho_A).
\end{align}
The homology constraint to the Ryu-Takayanagi formula, as depicted in Figure 2, implies that
there is a critical length $\ell_{{\rm crit}}$, 
where the Ryu-Takayanagi surface of $A$  becomes two disconnected pieces, 
 one wrapping the black hole and one homologous to $\bar{A}$.  
Above $\ell_{{\rm crit}}$, the entanglement entropy is given by,
\begin{align}
  S_{{\rm thermal}}(\ell_A) = S_{{\rm BH}} + S_{{\rm thermal}}(\pi - \ell_A), ~~~ \text{if} \ \ell_{\rm crit} \leq \ell_A .
  \label{reflection}
  \end{align}

To compute the Holevo information $\chi(A) = S(\rho_A) - \sum_i p_i S(\rho_{i, A})$, we also need to know  the
entanglement entropy for microstates, $ S(\rho_{i, A})$. In CFT in $(1+1)$ dimensions, this has been computed in \cite{Hartman:2014}
in a set-up of our interest. We will therefore discuss the Holevo information in AdS$_3$/CFT$_2$ in the next section.
 
\section{Holevo Information in AdS$_3$/CFT$_2$}

The Holevo information $\chi(A)$ is the difference bewteen the entanglement entropy 
for the thermal state $S(\rho_A)$ and the average entanglement entropy of microstates
$\sum_i p_i S(\rho_{i,A})$.
The former can be computed using the Ryu-Takayanagi formula in the AdS-Schwarzschild geometry and expressed in terms of the
hyperbolic functions of $\ell_A$.
According to  \cite{Hartman:2014}, 
when the central charge is large, the light spectrum is sparse, and 
$\rho_i$ corresponds to a heavy state in CFT$_2$ dual to a black hole microstate 
in AdS$_3$,
its entanglement entropy $S(\rho_{i, A})$ is given by,
\begin{equation} 
    S(\rho_{i, A}) = \begin{cases}
     S_{{\rm thermal}}(\ell_A), &\text{if}\ \ell_A \leq \pi/2 \\
    S_{{\rm thermal}}(\pi - \ell_A), &\text{if}\  \pi/2 <\ell_A .
    \end{cases}
    \end{equation}
Combining these results, we obtain
  \begin{equation}
    \chi(\ell_A)=  
    \begin{cases}
      0, & \text{if}\ \ell_A < \pi/2 \\
     \log\left[\frac{\sinh(2\pi\ell_A/\beta)}{\sinh(2\pi(\pi-\ell_A)/\beta)}\right], & \text{if}\
     \pi/2\leq \ell_A\leq l_{{\rm crit}} \\
      S_{{\rm BH}}, & \text{if}\   \ell_{{\rm crit}} < \ell_A ,
    \end{cases}
  \end{equation}
  where
   \begin{align}
  \ell_{{\rm crit}} = \frac{\beta}{4\pi} \log \left( \frac{1 + e^{(2\pi)^2/\beta}}{2} \right).
  \label{criticalsize}
  \end{align}
  
  The behavior of $\chi(A)$ is depicted in Figure 1. 
  Until $A$ covers one half of the Cauchy surface, $\chi(A)$ is identically equal to zero. 
  This means that there is no mutual information between
black hole microstates and  $A$, and we cannot distinguish between
microstates at all by measurements on $A$. Between $\pi$ and $\ell_{{\rm crit}}$,
$\chi(A)$ grows monotonically.
  At $\ell_A = \ell_{{\rm crit}}$, the Holevo information achieves its maximum value $S_{{\rm Shannon}}= S_{{\rm BH}}$.
For $\ell_A \geq \ell_{{\rm crit}}$,  microstates become perfectly distinguishable by measurements on $A$.  
  
The critical size $\ell_{{\rm crit}}$ depends on the size of the black hole
and therefore on its temperature. From its explicit form given by (\ref{criticalsize}),
we can see $\ell_{{\rm crit}} \rightarrow \pi$ in the high temperature limit, $\beta \rightarrow 0$.
This can be explained by the fact that the thermal ensemble $\rho$
is maximally mixed in this limit and we need to observe the whole
CFT to be able to distinguish each microstate. 


One loop corrections to the entanglement entropy \cite{Barrella:2013wja} above the Hawking-Page phase transition are suppressed by powers of $N$, and moreover are small and continuous around $\ell_{{\rm crit}}$. In particular, since the kink at $\ell_{{\rm crit}}$ is an $O(N^2)$ effect,
one loop corrections would not modify its qualitative feature.


 \section{Holevo Information  in AdS$_{d+1}$/CFT$_d$}
 
In higher dimensions, we do not have an explicit expression for the entanglement entropy $S(\rho_{i, A})$ of each microstate. 
Fortunately, the behavior $S(\rho_A)$ for the thermal ensemble depicted in Figure 2, combined with entropy 
inequalities that hold for general quantum systems, puts stringent constraints on $\sum_i p_i S(\rho_{i, A})$. 

To simplify equations, let us denote, 
\begin{align}
    & S_{{\rm micro}}(\ell_A)  \equiv \sum_i p_i S(\rho_{i, A}), \nonumber \\
    & \chi(\ell_A)  \equiv S_{{\rm thermal}}(\ell_A) - S_{{\rm micro}}(\ell_A). 
\end{align}
The upper and lower bounds on the Holevo information, 
\begin{align}
  0 \leq \chi(\ell_A) \leq S_{BH}, 
  \end{align}
 can be expressed as the bounds on $S_{{\rm micro}}(\ell_A)$ as,
  \begin{align}
    S_{{\rm thermal}}(\ell_A) -  S_{{\rm BH}}  \leq S_{{\rm micro}}(\ell_A) \leq S_{{\rm thermal}}(\ell_A)
    \label{Holevoinequalities}
\end{align}
If $\ell_{\rm crit} \leq \ell_A$, we can apply (\ref{reflection}) to rewrite the first inequality of the above as,
\begin{align}
S_{{\rm thermal}}(\pi - \ell_A) \leq S_{{\rm micro}}(\ell_A) ~~~~ \text{if}  \ \ell_{\rm crit} \leq \ell_A.
\label{inequalityone}
\end{align}
Since each microstate $\rho_i$ is pure,  $S(\rho_{i,A}) = S(\rho_{i, \bar{A}})$ where $\bar{A} = S^{d-1} \backslash A$.  
Therefore,
its ensemble average $S_{{\rm micro}}(\ell_A) = \sum_i p_i S(\rho_{i, A})$ is also reflection symmetric, 
\begin{align}
S_{{\rm micro}}(\pi -\ell_A)= S_{{\rm micro}}(\ell_A).
\end{align}
Using this, we can write (\ref{inequalityone}) as,
\begin{align}
S_{{\rm thermal}}(\ell_A) \leq S_{{\rm micro}}(\ell_A) ~~~~ \text{if}  \ \ell_A \leq \pi - \ell_{\rm crit}.
\label{inequalitytwo}
\end{align}
This is the opposite of the second inequality in (\ref{Holevoinequalities}) that holds for any $\ell_A$. Combining (\ref{Holevoinequalities})
and (\ref{inequalitytwo}), we conclude
$S_{{\rm micro}}(\ell_A) =  S_{{\rm thermal}}(\ell_A)$ for $\ell_A$ below $\pi -  \ell_{\rm crit}$.
Using (\ref{reflection}), we can also equate $S_{{\rm micro}}(\ell_A)$ with $S_{{\rm thermal}}(\ell_A) - S_{BH} $
in $\ell_A$ above $\ell_{{\rm crit}}$.

To summarize, we have been able to determine  $S_{{\rm micro}}(\ell)$  completely for the following ranges of $\ell_A$:
\begin{equation}
  S_{{\rm micro}}(\ell_A) =
\begin{cases}
  S_{{\rm thermal}}(\ell_A)  & \text{if}\ \ell_A \leq \pi - \ell_{{\rm crit}}, \\
 S_{{\rm thermal}}(\ell_A) - S_{BH} & \text{if}\ \ell_{{\rm crit}} \leq \ell_A.
\end{cases}
\label{sameasblackhole}
\end{equation}
Therefore, the Holevo information in these ranges is given by, 
\begin{equation}
  \chi(\ell_A) =
\begin{cases}
 0  & \text{if}\ \ell_A \leq \pi - \ell_{{\rm crit}},  \\
 S_{BH} & \text{if}\ \ell_{{\rm crit}} \leq \ell_A.
\end{cases}
\end{equation}
As in the case of AdS$_3$/CFT$_2$, microstates are not distinguishable at all for $\ell_A \leq \pi - \ell_{{\rm crit}}$
and are perfectly distinguishable for  $\ell_{{\rm crit}} \leq \ell_A$.

Less is known about $S(\rho_{i, A})$ in the range  $\pi - \ell_{{\rm crit}} < \ell_A < \ell_{{\rm crit}}$ for general $d$.
Since $S(\rho_{i, A})$ is not an expectation value of
an observable, it is not clear if the eigenstate thermalization hypothesis implies that
the Ryu-Takayanagi formula can be used to compute this quantity.
 In the AdS-Schwarzschild solution, 
 there is always a minimal surface that stays outside of the
 horizon \cite{Hubeny}. However, a microstate may have geometry 
 inside of the horizon that is different from that of the AdS-Schwarzschild solution,
allowing a minimal surface take a shortcut and making $S(\rho_{i, A})$ smaller. 
 It is also possible that quantum or stringy effects become enhanced near or inside of the horizon,
making large corrections to $S(\rho_{i, A})$.
Our result (\ref{sameasblackhole}) suggests
that such corrections do not take place  
in the ranges $\ell_A \leq \pi - \ell_{{\rm crit}}$ and $\ell_{{\rm crit}} \leq \ell_A $, perhaps because
 minimal surfaces subtending $A$ stay sufficiently far away from the horizon.
 
A possible argument against the possibility of shortcuts across the horizon region was given in \cite{Engelhardt:2013tra}.  Suppose there is such a shortcut. We may then consider reducing the size of the boundary domain. In this case, the extremal surface would eventually touch the horizon. But, according to \cite{Engelhardt:2013tra}, if an extremal surface is ever even pointwise tangent to the horizon it must wrap around it, thus precluding the possibility of a shortcut\footnote{We thank Netta Engelhardt for informing us of this argument.}.

For general $d$, we can show that such corrections can only 
increase the Holevo information $\chi(A)$.
Using the definition (\ref{Holevo}) of $\chi(A)$  
and the reflection symmetry $S_{\rm micro}(\pi - \ell_A) =
S_{\rm micro}(\ell_A)$, we can express
the Holevo information as, 
\begin{align}
   \chi(A) = S(\rho_A) - S(\rho_{\bar{A}}) + \chi(\bar{A}).
 \label{anotherexpression}
\end{align}
and the ensemble average of  $S(\rho_{i, A})$ as,
\begin{align}
 \sum_i p_i S(\rho_{i, A}) =  S(\rho_{\bar{A}}) - \chi(\bar{A}). 
 \label{entanglementcorrections}
 \end{align}
Since $\chi(\bar{A}) \geq 0$, these quantities are bounded as,
\begin{align}
\chi(A) \geq  S(\rho_A) - S(\rho_{\bar{A}}), ~~~
 \sum_i p_i S(\rho_{i, A}) \leq  S(\rho_{\bar{A}}),
 \end{align}
 for the entire range of $\ell_A$ for general $d$.
 In particular, for $\ell_A \geq \pi/2$, the ensemble average of  $S(\rho_{i, A})$
 is bounded above by the naive application of the Ryu-Takayanagi formula to these microstates,
 namely any corrections to the minimal surface calculation would decrease the entanglement entropy.

\section{Discussion}

We found that black hole microstates become
perfectly distinguishable at $\ell_A = \ell_{{\rm crit}}$,
where the Holevo information achieves its maximum value.
 What information are we missing in $\ell_A < \ell_{{\rm crit}}$? 
 One may have thought that the missing information is carried by  
 $\bar{A}$,
 the complement of $A$. 
 However, 
 the mutual information between microstates and measurements   
on $\bar{A}$ is bounded above by $\chi(\bar{A})$,
 which in general is different from the lower bound 
 $(S_{{\rm BH}} - \chi(A))$ of the missing information.
In particular, since $\chi(\bar{A})=0$ for $\ell_A\geq \pi/2$ in AdS$_3$/CFT$_2$ as shown Figure 1, measurements on $\bar{A}$
alone would not provide any information on microstates in this case.

A useful expression for $(S_{{\rm BH}}-\chi(A))$
can be obtained by using a purification of $\rho$,
\begin{align}
| \Psi \rangle = \sum_i \sqrt{p_i} \ |\psi_i\rangle\otimes |\psi'_i \rangle,
\end{align}  
where $|\psi'_i\rangle$'s are orthonormal states in the purifying Hilbert space ${\cal H}'$,
which we regard as a copy of the CFT Hilbert space ${\cal H}$. 
For the eternal black hole, $|\Psi\rangle$ is the thermofield double state expressing the entanglement through the 
Einstein-Rosen bridge of the black hole \cite{Maldacena:2001kr}.
The mutual information $I({\cal H}': \bar{A})$ between 
${\cal H}'$ and $\bar{A}$ for $| \Psi \rangle$ is given by,
\begin{align}
  I({\cal H}': \bar{A}) &=
 S(\rho_{{\cal H}'}) + S(\rho_{\bar{A}}) - S(\rho_{{\cal H}'\cup \bar{A}}) \nonumber \\
 &=   S_{{\rm BH}} +  S(\rho_{\bar{A}}) - S(\rho_A),
 \label{pure}
  \end{align}
where we used the fact that $A$ is the purification of $\rho_{{\cal H}'\cup \bar{A}}$.
Combining (\ref{pure}) and (\ref{anotherexpression}), we can express the lower bound on the missing information as,  
   \begin{align}
   S_{{\rm BH}} -\chi(A) =  I({\cal H}':\bar{A}) - \chi(\bar{A}). 
   \label{purification}
   \end{align}

In particular, in AdS$_3$/CFT$_2$, 
\begin{align}
\chi(A) = S(\rho_A)- S(\rho_{\bar{A}}), ~~~ S_{{\rm BH}} - \chi(A)  = I({\cal H}':\bar{A}),
\end{align} 
for $\ell_A\geq \pi/2$  since $\chi(\bar{A})=0$. We can interpret these using the bit threads picture of 
the Ryu-Takayanagi formula \cite{Freedman:2016zud} as follows.
In this picture, the minimal surface
serves as a constraint surface upon which the bit threads used to calculate the entropy of the subtended boundary region 
are evenly spaced, thus enforcing that the number of bit threads reproduces the Ryu-Takayanagi entopy. For the black hole
thermal ensemble in AdS$_3$/CFT$_2$, $\chi(A) = S(\rho_A)-S(\rho_{\bar{A}})>0$ for 
$\ell_A > \pi/2$ due to the homology constraint. 
This means that the number of bit threads that cross the Ryu-Takayangi surface subtending $A$ will be larger than the number 
of bit threads that cross the surface subtending $\bar{A}$. 
These excess threads therefore begin to attach on the black hole event horizon. At $\ell_A=\ell_{{\rm crit}}$, the black hole surface becomes saturated by bit threads, given their even spacing requirement.
Above $\ell_{{\rm crit}}$, all threads with one end on the black 
hole event horizon must end on $A$. 
Therefore, if we interpret the Holevo information as the
number of threads connecting $A$ and the event horizon, 
it reproduces its behavior. 

However, the situation may become more complicated in higher dimensions.
As we pointed out at the end of the previous section, 
the ensemble average of the microstate entanglement entropy 
may be smaller than the one expected from 
the naive application of the Ryu-Takayanagi formula 
due to a shortcut in microstate geometry or enhanced 
quantum/stringy effects near or inside of the horizon.
Such corrections, if any, are captured by $\chi({\bar A})$ as shown in 
(\ref{entanglementcorrections}). It would be interesting if they can also be
accounted for by the bit threads picture.

According to the entanglement wedge reconstruction picture, 
the subregion $A$ of the CFT carries information necessary to reconstruct
local excitations in the bulk region bounded by $A$ and the Ryu-Takayanagi surface subtending $A$. 
Since the naive application of the Ryu-Takayanagi
formula to a microstate 
suggests that the entanglement wedge  for $\pi/2 \leq\ell_A\leq\ell_{{\rm crit}}$ 
covers the entire black hole region, one may have expected that
maximum distinguishablity of microstates should already have been achieved in this range of $\ell_A$,
contradicting our results shown in Figure 1.
However, the entanglement wedge for $A$ can depend on the choice of the code subspace \cite{Harlow:2016vwg}.
If we choose the code subspace to be too small, it may not carry enough information to 
distinguish microstates. Our results suggest that, if  the code subspace is large enough to
differentiate one microstate from another, the entanglement wedge for $A$ covers
the black hole region only when $\ell_A \geq \ell_{{\rm crit}}$. 

It may also be possible to demonstrate this using tensor network models. Since a black hole
in a tensor network 
is simply the deletion of tensors in the center of the bulk, one naturally reproduces the Ryu-Takayanagi surfaces that avoid the event horizon
and the entanglement entropy $S(\rho_A)$ for the mixed state $\rho$.
It has been conjectured in \cite{Harlow:2016vwg} that there may exist other choices of the code subspace that would allow the
entanglement entropy $S(\rho_{i, A})$ for a pure state $\rho_i$ 
to be given by a Ryu-Takayanagi surface penetrating 
the would-be horizon region. This would allow measurements on $A$
to have access to the interior of the microstate and explain
the gradual rise of $\chi(A)$ in $\pi - \ell_{{\rm crit}} < \ell_A <
\ell_{{\rm crit}}$.

\begin{figure}[t]
  \begin{center}
    \includegraphics[width=0.6\textwidth]{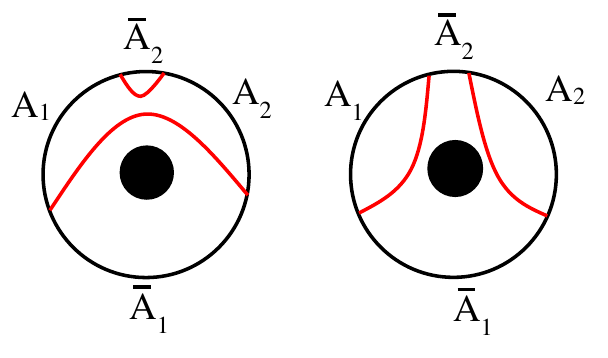}
\end{center}
  \caption{Two possible Ryu-Takayanagi configurations for $A_1 \cup A_2$. The total area of the surfaces on the left is $S(\rho_{A_1\cup A_2})=S(\rho_{\bar{A}_2})+S(\rho_{A_1\cup \bar{A}_2 \cup A_2})$, while 
  that on the right is $S(\rho_{A_1 \cup A_2})=S(\rho_{A_1})+S(\rho_{A_2})$.}
\end{figure}

So far, we have considered the case when $A$ is connected. 
It is possible to decrease the total size of $A$ necessary for perfect distinguishability
by allowing $A$ to be disconnected \cite{Pastawski:2016qrs}. Let us consider $A$ for  $\ell_A < \ell_{{\rm crit}}$. 
Suppose we remove from $A$ a region antipodally located to the center of $\bar{A}$, label this new region $\bar{A}_2$,
 and relabel the original $\bar{A}$ to be $\bar{A}_1$. Furthermore, label the two now disconnected components of $A$ as $A_1$ and $A_2$. A pictorial depiction of this is in Figure 3.
So long as $S(\rho_{A_1\cup \bar{A}_2 \cup A_2})+S(\rho_{\bar{A}_2})\leq S(\rho_{A_1})+S(\rho_{A_2})$, the addition of the $\bar{A}_2$ Ryu-Takayanagi surface to the original RT surfaces for $A$ will be the minimal surface homologous to $A_1\cup A_2$. Thus, the
phase transition still happens at $\ell_{{\rm crit}}$, but the total size of $A$ is decreased by the
size of $\bar{A}_2$, ensuring perfect distinguishability with a smaller fraction of the CFT boundary than the one interval case.

Below the Hawking-Page phase transition, one would expect the Holevo information to increase monotonically from zero boundary region size. It is possible that it can be computed directly using the one-loop results of \cite{Barrella:2013wja}, as the leading-order Ryu-Takayanagi pieces should cancel, though some care must be taken as to how to calculate the entanglement entropy of microstates of the thermal AdS ensemble. For example, the one-loop contribution to the entanglement entropy of $A$ given in \cite{Barrella:2013wja} would not be the same as that of $A^c$, something which should be true for pure microstates.

\section*{Acknowledgement}

We would like to thank Aidan Chatwin-Davies, Xi Dong, Netta Engelhardt, Daniel Harlow, Thomas Hartman,  Akio Hosoya,
Veronika Hubeny,
 Tomonori Ugajin, and Beni Yoshida  for discussions.
This research is supported in part by
U.S.\ Department of Energy grant DE-SC0011632.
N.B. is also supported in part by the DuBridge Fellowship of the Walter Burke Institute for Theoretical Physics. 
H.O. is also supported in part
by JSPS Grant-in-Aid for Scientific Research C-26400240 and 15H05895.
The Kavli Institute for the Physics and Mathematics of the Universe is
supported in part  
by the World Premier International Research Center Initiative,
MEXT, Japan.

\end{document}